\documentclass[a4paper,12pt,oneside]{article}
\usepackage{graphicx}
\usepackage{amsmath,amssymb}			

\begin{document}

\title{The jet of the young star RW~Aur\,A and related problems}

\author{L.~N.~Berdnikov$^{1,2}$, M.~A.~Burlak$^1,$ O.~V.~Vozyakova$^1,$\\
A.~V.~Dodin$^1,$ S.A.~Lamzin$^1$ 
\footnote{E-mail: lamzin@sai.msu.ru},
and A.~M.~Tatarnikov$^1,$}

\date{
$^1$ Sternberg Astronomical Institute of Lomonosov Moscow State University,
Russia \\
$^2$ Astronomy and Astrophysics Research division, Entoto
Observatory and Research Center, Addis Ababa, Ethiopia \\
}

\maketitle

\medskip


\section*{Abstract}
Comparing the images of the jet of the young star RW~Aur\,A, separated by a period of 21.3 years, we found that the outermost jet's knots have emerged $\approx 350$ yr ago.  We argue that at that moment the jet itself has appeared and intensive accretion onto the star has began due to the rearrangement of its protoplanetary disk structure caused by the tidal effect of the companion RW~Aur\,B. More precisely, we assume that the increase of accretion is a response to changing conditions in the outer disk regions, which followed after the sound wave, generated by these changes, crossed the disk in a radial direction.  The difference in the parameters of the blue and red lobes of the RW~Aur\,A jet, according to our opinion, is a result of the asymmetric distribution of the circumstellar matter above and below the disk, due to a fly-by of the companion.  It was found from the analysis of RW~Aur historical light curve that deep and long $(\Delta t>150$ days) dimmings of RW~Aur\,A observed after 2010 yr, had no analogues in the previous 110 years.  We also associate the change in the character of the photometric variability of the star with the rearrangement of the structure of inner $(r<1$ a.u.) regions of its protoplanetary disk and discuss why these changes began only 350 years after the beginning of the active accretion phase.
%

%
\section*{Introduction}

   RW~Aur\,A is a classical T Tauri star (CTTS). It means that it is a young
low mass pre-main-sequence star, activity of which (variable brightness,
excess emission in lines and continuum in all spectral bands, powerful
outflow from its vicinity etc.) in some way connected with accretion of
protoplanetary disk matter \cite{HHC-16}.  It has a companion \cite{JvB-44},
which is now at angular distance $\approx 1.5^{\prime\prime}$
\cite{Bisiklo-12,Cs-17}, what corrsponds to projected separation 210 a.u., 
if to adopt that the distance to RW~Aur is 140 pc \cite{Elias-78}.

  RW~Aur\,A is one of the most active CTTSs what seems quite strange on
closer examination. Indeed, estimations of the accretion rate from the
protoplanetary disk onto the star $\dot M_{ac}$ yield a value greater than
$2\times 10^{-8}$ $M_\odot$/yr [6,7], while the mass and radius of the
stellar protoplanetary disk are relatively small: $M_d=4\times10^{-3}$
$M_\odot$ \cite{AW-05} and $R_d < 80$ AU \cite{Cabrit-06}.  Therefore, the
characteristic time of being in the active state for RW~Aur\,A, which is $t_d \sim
M_d/\dot M_{ac} \lesssim 2\times 10^4$ yr, appears much less than its
age, which knowingly exceeds 1 Myr \cite{White-Ghez-01}.

  To explain this fact the authors of \cite{Cabrit-06} supposed that the
observed high accretion rate was due to tidal disturbance of the RW~Aur\,A
disk caused by fly-by of another star -- RW Aur B. Simulations carried out
by \cite{Dai-15} showed that the observed structure of outer regions of the
RW~Aur\,A disk could be reproduced if a companion fly-by the primary at a
minimum distance of about 70 AU 400-450 yr ago.

  The fact that passage of the companion resulted in the change of dynamics of
inner parts of the RW Aur A disk was also discussed after unexpected long-term
($\Delta t \sim 150^d)$ and quite deep $(\Delta V \sim 2.5^m$) dimming
occured in 2010-2011 \cite{Rodriguez-etal-13}, which repeated three years
later on a greater scale: $\Delta t\approx 2$ yr, $\Delta V > 5^m$
\cite{Lamzin-16}.  Possibility of intensification of a "dusty"\, wind from
inner $(r < 1$ AU) regions of the disk is considered as a probable cause of
these events \cite{Petrov-15,Shenavrin-15, Bozhinova-16,Facchini-16}.

It is well-known that there is an intensive outflow of matter from the
RW~Aur\,A (see, e.g., \cite{Petrov-01,Alencar-05} and references therein). 
\cite{Hirth-94} have found a bipolar jet in the star, which is a collimated
part of the outflow.  According to common terminology, the part of the jet
moving toward the Earth will be called the blue lobe of the jet and the part
moving away from the Earth -- the red lobe.  The radial velocity and
physical parameters of gas in the blue and red lobes prove to be
considerably different as well as their spatial extension (see e.g. 
\cite{Melnikov-09} and references therein).  Each jet consists of a chain of
knots\footnote{
In fact each knot is a Herbig-Haro object, but in the catalog of these
objects \cite{Reipurth-99}, the reference number HH\,229 has been assigned
to the whole complex of the knots.
},
the number of which are different in the lobes. The authors
\cite{Lopez-Martin-03} have found that knots closest to the star
$(d<5^{\prime\prime})$ move away from it graduall changing their shapes.

  It is generally thought that emerging of knots in jets is caused by
non-stationary processes in the wind-formation region -- stellar
magnetosphere and/or the innermost regions of its accretion disk -- during
which short-term increasing of gas velocity takes place (see
\cite{Krist-08,Frank-14} and references therein).

  The aim of the present work is to associate the changes in the morphology
of the RW~Aur\,A jet, which happened in two past decades, and in the
photometric behavior of the star after 2010 with tidal influence of the
companion on the inner disk regions of the primary.


\section*{Observations}

  Observations of RW~Aur were carried out with the 2.5-m telescope of the
Caucasian Mountain Observatory of SAI MSU in 2017.  In the optical range
images of star's neighborhood of a size of $10^{\prime}\times10^{\prime}$
were obtained with the [S\,II] filter $(\lambda_c=672.17$ nm, $FWHM= 5.68$
nm) on February 20 and 24 and with the $H_\alpha$ filter $(\lambda_c=656$
nm, $FWHM= 7.67$ nm) on February 28 with the total exposure time 2500 and
2100 s respectively.  The seeing varied from 0.8 to 1.3${}^{\prime\prime}.$
We used E2V CCD44-82 CCD with a pixel size of 15 $\mu$m as the light detector.
The images obtained were corrected for BIAS and flat field in the
standard way.  Using coordinates of more than 50 field stars taken from the
USNO-B1.0 catalog \cite{Monet-03}, we found that the image scale is 0.1543
${}^{\prime\prime}$/pixel.

  On February 24 and 28, 2017, we also obtained images of the region around
RW~Aur in the [Fe\,II] band ($(\lambda_c=1644.2$ nm, bandwidth -- 26.2 nm) in
the photometric mode of the ASTRONIRCAM infrared camera-spectrograph
\cite{Nadzhip-17}. The image scale was 0.2695${}^{\prime\prime}$/pixel, the
field of view -- $4.6^\prime$. The resulting image is a sum of 80 images
obtained on February 24 with the accumulation time equal to 25 s and image
quality $\approx 1.3^{\prime\prime};$ and a sum of 180 images with the
accumulation time equal to 16 s and image quality $\approx
1.0^{\prime\prime},$ obtained on February 28. Thus, the total exposure time
exceeded 80 minutes. 

  The resulting images in the [S\,II], $H_\alpha$ , and [Fe\,II] bands are
presented in three bottom panels of Fig.~1.


\section*{Proper motion of the jet's knots}

 Observations \cite{Hirth-94}, from which the jet of RW~Aur\,A was found,
covered the $30^{\prime\prime}$ region on both sides of the star.  In
October 1995, the authors of \cite{Mund-Eisloffel-98} obtained a much larger
image of the neighborhood of the star in the [S\,II] band close in
parameters to that we used.  They have found that spots are seen up to
$\approx 120^{\prime\prime}$ (the $I$ knot) in the blue lobe and up
to $\approx 55^{\prime\prime}$ in the red lobe (the $M$ knot) (see the top
panel of Fig.\ref{Fig-1}, which is a copy of Fig.~2 from
\cite{Mund-Eisloffel-98}).
  
  Images of inner $(d<5^{\prime\prime})$  jet regions in the [S\,II] lines
with high angular resolution were obtained in \cite{Dougados-00, Woitas-02}
on December 30, 1997 and December 10, 2000 respectively.  Having compared
these images and those obtained by themselves on December 19, 1998, the
authors of \cite{Lopez-Martin-03} determined the proper motion velocity of
the knots in the inner region of the jet and inferred that the knots are
moving away from the star 1.6 times faster in the blue lobe than in the red
one.

  In order to derive the tangential velocity of the knot motion $\mu$  at a
greater distance from the star and in a longer time interval, we compared
the image we obtained in the [S\,II] band and the data on the distance from
the spots to the star $d$ in the similar band in different epochs: the values
for $d_{1997}$ and $d_{2000}$ were taken from Table~1 of
\cite{Lopez-Martin-03}, and the values for $d_{1995}$ were obtained from
Fig.~2 in \cite{Mund-Eisloffel-98}, which we processed in the same way as
images we observed.  Our images in the $H_\alpha$ and [Fe\,II] $\lambda=1.64$
$\mu$m bands were not used to determine $\mu$, as far as positions of the
knots in the images slightly differ in various bands.  For each knot, Table
1 gives distances from the point of its maximum intensity to RW~Aur\,A
(except for the $G$ knot), the $\mu$ value calculated from the difference of
coordinates, and also the "age"\, of the knot $t_d=d/\mu.$ The knot $G$ is
crescent-shaped and we considered its position to be best characterized by
the coordinate of the point most distant from the star at the arching edge
of a crescent.  Note that the knot marked in the table as $R_?$ has no
designation in the paper \cite{Lopez-Martin-03}.

The details inside the region of the radius $5-10^{\prime\prime}$ around
RW~Aur are indistinguishable in Figure~1. However, $A$ and $B$ components
of the binary system, the angular distance between which is $1.5^{\prime\prime},$
are confidently separate in all our images, so that we have quite reliably
determined the centroids of the knots close to RW~Aur\,A after subtracting
the wings of images of both components.

 It follows from the Table~1 that the mean proper motion $\mu$ of six knots
in the red lobe for about past 20 yr is equal to $0.18 \pm 0.03$
${}^{\prime\prime}$/yr.  This coincides, within error, with the value of
$0.16$ ${}^{\prime\prime}$/yr found by \cite{Lopez-Martin-03} from two knots
in the time interval of about two years.  In the blue lobe, these authors
measured $\mu$ for the $B_3$ knot only.  They found $0.26 \pm 0.035$
${}^{\prime\prime}$/yr, which is also in agreement with the value we
obtained $0.26 \pm 0.02$ ${}^{\prime\prime}$/yr.  However, we found a
slightly greater mean value $\mu$ for the knots $B,$ $C_1,$ $C_2,$ $D,$ and
$G$ of the blue lobe: $0.34 \pm 0.02$ ${}^{\prime\prime}$/yr.  It is
difficult to say how significant this difference is, although, if one
calculates the mean proper motion from all knots of the jet blue lobe, it
will be equal to $0.33 \pm 0.04$ ${}^{\prime\prime}$/yr, i.e.  1.8 times
greater than that of the red lobe, which, within error, coincides with the
ratio found in \cite{Lopez-Martin-03}.\footnote{
Mean $\mu$ values correspond to motion of knots perpendicularly
to the line of sight with a linear velocity of about 160 km s$^{-1}$ in the
blue lobe and about 90 km s$^{-1}$ in the red lobe.
}
%


%
\section*{Estimation of the jet's age}

  If to divide the coordinates of the outermost knots $M$ and $I$ in the
jet's 1995 image by the mean $\mu$ values in the corresponding lobes and to
add 21.3 yr, one can found that they have appeared $320 \pm 50$ and $380 \pm
40$ yr ago respectively.

  A number of the knots seen in the 1995 image are absent in the 2017 image. 
This is the case, in particular, of the bright knot $H$ and also the knots
most distant from the star -- $I$ and $M.$ On the other hand, there are
spots in the 2017 image marked as $N_1-N_3$ that have not been noticed
earlier.  Thus, one cannot exclude that there had been knots before 1995
which were farther from the star than the knots $I$ and $M,$ but they have
disappeared by 1995.  Formally, this means that the age of the jet we
determined, $t_{jet} \approx 350$ yr, is the lower limit.  However, as can
be seen from Fig.~4 in \cite{Dai-15}, formation of tidal arms in the disk of
A-component has started not earlier than 200 years before the periastron
passage of the companion, i.e., approximately less than 650 years ago, so
our value $t_{jet}$ should be close to realistic. Moreover, increasing the
age of the jet means reducing the time $t_{exc}$ that necessary for tidal
disturbance to come from the outer regions to the innermost, while the value
$t_{exc}$ we obtained is already quite short: $650-t_{jet} \approx 300$ yr.

  Let us clarify the foregoing. As can be seen from the same figure, during
fly-by of the companion the outer regions of the RW~Aur\,A disk undergo
strong deformation and, in fact, are transformed into two spiral arms, one
of which stretches toward the companion and the other -- in the opposite
direction.  However, even at a minimal approach of the stars, the influence
of the companion on the innermost regions of the A-component disk is
extremely small.  Consequently, the burst in the accretion activity of
RW~Aur\,A is a response not to the gravitational field of the companion, but
to a change in physical conditions at the outer boundary of the disk, which
manifests itself after some time $t_{exc}.$

  But how is the relevant information passed from the outer regions of
the disk to the inner ones?  In calculations of the dynamics of the RW~Aur\,A
disk under the influence of the companion's fly-by, the authors of
\cite{Dai-15} considered only the regions distant from the primary star over
6 AU.  Since we are interested in much closer neighborhood, we are forced to
confine ourselves to corresponding estimates.  According to
\cite{Abolmasov-16} the characteristic time for the rearrangement of the
radial structure of the disk of RW~Aur\,A (the so-called "viscous time"\,)
is
\begin{equation}
t_{vis}={t_k \over 2\pi \alpha} {\left( {H\over R_d} \right)}^{-2}, 
\quad
t_k = 2\pi {\left( {R_d^3 \over GM} \right)}^{1/2}.
\label{eq:1}
\end{equation}

  Here $M=1.4\,M_\odot$ \cite{Woitas-01} is the mass of the star; $R_d
\approx 60$ AU \cite{Dai-15} is the external radius of the disk before the
companion's fly-by; $\alpha$ and $H$ -- are the Shakura-Sunyaev parameter
and the half-thickness of the disk at the outer boundary, respectively.  The
values $\alpha \sim 0.01$ and $H/R_d \sim 0.1$ are considered as typical of
the CTTS's disks \cite{HHC-16}, from where it follows that the orbital
period at the outer disk boundary ("Keplerian time"\,) $t_k \approx 400$ yr,
and $t_{vis} \approx 6\cdot10^5$ yr. Thus, disturbances that led to the
birth of the jet transited from the periphery to inner regions of the disk
in time that much less than $t_{vis},$ but comparable with the Keplerian
time.

  Let us now estimate how much time $t_{hyd}$ it will take the sound wave to
cross the disk "to inform"\, its inner regions about pressure variations at
the outer boundary.  According to \cite{Cabrit-06}, the temperature at the
outer boundary of RW~Aur\,A's undisturbed disk is $\approx 30$ K, and at
present epoch, the gas temperature (which is already heated by the tidal
interaction) is $80 \pm 20$ K.  Considering that gas of outer disk regions
mainly consists of molecular hydrogen, we obtain that the sound speed $V_s$
in this region is 0.6 km s$^{-1}$ with accuracy of $\sim 30$\,\%. 
Consequently, $t_{hyd}\sim R_d/V_s \sim 500$ yr.  \footnote{
It is not by chance that we obtained almost similar values of times $t_k$
and $t_{hyd}.$ If to write down the expressions for the scale hight of an
$\alpha$-disk at its outer boundary as $H=V_s t_k/2\pi$ \cite{Abolmasov-16},
then one obtain $t_k/t_{hyd}=2\pi \left(H/R_d \right) \sim 1,$ because
$H/R_d \sim 0.1$ in our case.
}

  The observed difference of velocities of the companion and the primary
star is about of several km s$^{-1}$ \cite{White-Hillebrandt-04,Bisiklo-12}. 
Therefore, the relative velocity of the stars, when they have approached
each other, was about an order of magnitude higher than $V_s.$ If it follows
from this fact that the disturbance propagated through the disk with
supersonic velocity, at least in its outermost regions, then the time
$t_{exc}$ is in a full agreement with the restriction resulting from our
observations: $\lesssim 300$ yr.


%
\section*{Analysis of RW~Aur's historical lightcurve}

 However, the question on how the passage of the companion affected the
inner regions of the disk of RW~Aur\,A is not exhausted by this coincidence. 
We have already mentioned in the Introduction that in the current decade
several long-lasting and deep dimmings of the A-component have occured, which
were apparently associated with the rearrangement of the structure of the
outer disk regions in some way or other.  To understand how unique these
events are, let us consider the historical light curve of RW~Aur.

  Since L.~P. Ceraski detected the brightness variability of this star
\cite{Cerasky-1906}, the character and possible causes of these variations
at the time intervals from minutes to several years were repeatedly
discussed (see, e.g., \cite{Tsesevich-Dragomiretskaya-73,
Chugainov-Lovkaya-88, Chugainov-90, Herbst-etal-94, Petrov-15,
Beck-Simon-01, Grankin-etal-07, Rodriguez-etal-13, Rodriguez-etal-16,
Bozhinova-16} and references therein).  Strictly speaking in these cases
summary brightness of A and B components was used, as the distance between
them is $<1.5^{\prime\prime}.$ However, resolved photometry
presented in \cite{White-Ghez-01,Antipin-etal-15} showed that until 2010 the
companion was fainter than A-component by $2-2.5$ mag in the $B$ and $V$
bands, so it can be accepted that the integrated brightness variations
reflect the behavior of RW~Aur\,A.  

  The historical light curve presented in Fig.~2 was constructed using
photoelectric, photographic, and visual observations on the assumption that
photographic and visual measurements are identical to the photoelectric in
the $B$ and $V$ bands, respectively.  Photoelectric data were taken from
\cite{Herbst-etal-94}, the RОТОR \cite{Grankin-etal-07}, and AAVSO
\cite{Kafka-15} databases as well as from papers \cite{Babina-etal-13,
Rodriguez-etal-13}.

  Visual estimates are taken from the AAVSO database and 3287 photographic
estimates of brightness -- from the book
\cite{Tsesevich-Dragomiretskaya-73}, such as 2748 of them are based on
measurements of plates from the Harvard College Observatory (USA)
Astronomical Plate Stacks observed in the period of JD
2\,414\,639--2\,437\,288.  Additionally, we estimated the brightness of
RW~Aur from 166 photo plates from the same collection in the time interval
of JD 2\,439\,801--2\,447\,862.  We studied the DNB photo plates acquired
using a 38 mm aperture camera; they have a resolution of
580~${}^{\prime\prime}/$mm and a limiting magnitude of about $15^m.$ Let us
note that the historical light curve given in Fig.~3 in \cite{Beck-Simon-01}
was constructed from the measurements of 150 photoplates from the same
collection.  In the digital form, these estimates are not presented in the
paper, but most likely they are among our much more numerous data.

 Let us consider initially the period before 2010. Figure 2 shows that at
that time the brightness of the star during the seasons varied without any
noticeable regularity, although, single episodes of about several tens of
days took place in which the stellar brightness attenuated and then reverted
to the initial level: examples of such episodes are presented in Fig.~3.

  According to photoelectric measurements only, started from the beginning
of the 60s, the average brightness of RW~Aur in the $B$ and $V$ bands is
equal to $11.2$ and $10.5$ mag, respectively.  The amplitude of season
variations in the V band was about 1.4 mag, such as the brightness of the
star during the whole period of photoelectric observations never dropped
below $11.8^m.$ Among $\approx 10\,000$ visual estimates of brightness
before 2010, the lower values (down to $V = 12.6$ mag) were observed in
twenty cases only, however, some of these estimates were undervalued and in
other cases duration of a state with $V > 12^m$ usually did not exceed
several days (see, e.g., the central panel of Fig.~3).  The star
demonstrated similar behavior but with a greater variability amplitude in
the $B$ band: as can be seen from the top panel of the same figure, the
stellar brightness in this range could vary by more than $2.5^m,$ for about
a month, which was also mentioned in \cite{Beck-Simon-01}.

  It also follows from Figure 2 that during the past century mean seasonal
values of brightness have undergone wavy, aperiodic variations.  Note that
decreasing in the mean seasonal level of brightness at the end of the 30s to
the minimum value apparently reflects real changes happened to the star,
because the minimum is observed both in the $B$ and $V$ bands.  Thus, the
photometric behavior of RW~Aur in the period under study at all time scales
was nonstationary and non-trivial, probably because the brightness
variability was due to the combined effect of two mechanisms: non-stationary
accretion and shielding of starlight by relatively small gas-dust clouds
\cite{Petrov-Kozak-07}.

  However, even on this background, the stellar brightness attenuation in
2010-2011 lasted for about 200 days, during which $V$ magnitude was weaker
$13^m$ \cite{Rodriguez-etal-13}, became an extraordinary event.  Three years
later, the dimming of even greater scale has occured: it lasted for two
years and the brightness of RW~Aur\,A fell down to $V\approx 15.1^m$ at the
minimum \cite{Lamzin-16}.  The intensive study of the star in this period in
the range from 5 $\mu$m to 10 keV showed that attenuation of the brightness
of A-component happened due to its eclipse by a gas-dust cloud distant from
the star by less than 1 AU \cite{Petrov-15, Shenavrin-15, Schneider-etal-15,
Bozhinova-16, Rodriguez-etal-13, Rodriguez-etal-16, Takami-etal-16,
Facchini-16}.  By August 2016, the brightness of RW~Aur has returned to the
average level before 2010, but two months later a new eclipse has began,
which is lasting till now (April 2017).

 It is assumed that the disk of RW~Aur\,A is inclined to the line of sight
at an angle of $30-45^\circ$ \cite{Cabrit-06}, so the issue about what made
the dust to rise so high and stay so long above the disk plane is actively
discussed in the above-mentioned papers.  Intensification of "dusty wind"\,
from the inner $(r < 1$ AU) disk regions and/or deformation (bending) of
these regions are considered as a probable reason for this.  However, even
if deformation of the disk is associated with an existence of close low-mass
companion \cite{Facchini-16}, the question arises: why have not such deep
and long-term eclipses taken place before 2010?

The same question can be formulated in another way.  Whatever the mechanism
for the appearance of large-scale dust clouds in the line of sight after
2010 is, its starting as well as formation of the jet is associated with the
rearrangement of the inner disk regions after the fly-by of the companion
near RW~Aur\,A.  The question then is: why are these two events separated by a
time span of about 350 yr?

  It can be assumed that the scale and/or nature of the processes causing
the appearance of dust in the line of sight have changed after 2010 due to
the fact that an outwardly propagating wave of adjustment of the outer disk
regions to the changed conditions at the inner boundary has reached the
corresponding region.  If we put $t_{vis}=350$ yr, then it follows from
relation (1) that during this time the rearrangement of the radial structure
of the disk can occur up to distances of about 0.4 AU.  It is intersting
that, apparently, the dust cloud eclipsing the star after 2010 was at
approximately the same distance from RW~Aur\,A \cite{Shenavrin-15}.  We
cannot say whether such a coincidence is accidental.


%
\section*{On a cause of asymmetry of RW~Aur\,A jet}

 Consider now the problem of asymmetry of RW~Aur\,A jet, which is manifested
in the difference in the morphology, physical parameters of the gas
(density, temperature, and ionization degree), as well as velocities of the
knots in the blue and red lobes.  Similar assymetry is quite often observed
in CTTS jets, and, in principle, can be due to an asymmetry of the
parameters of the circumstellar environment or/and the region in which the jet
is formed (see, for example, \cite{White-etal-14} and referenced therein). 
The authors of \cite{Melnikov-09} found that the mass loss rates in the red
and blue lobes of the RW~Aur\,A jet are approximately equal $(2.6\times
10^{-9}$ and $2.0\times 10^{-9}$ $M_\odot/$yr, respectively) and therefore
the reason for jet's asymmetry is associated with the difference in the
properties of the environment.

  According to the results of calculations \cite{Dai-15}, the companion
moves around RW~Aur\,A in an orbit close to parabolic, the plane of which is
inclined to the plane of the disk $(z=0)$ at an angle of $18^\circ.$
Initially $(t=-\infty)$, the companion was in that region of space
(relatively speaking at $z>0),$ where the blue lobe of the jet was located,
moving from top to bottom.  At some moment the companion crossed the plane
$z=0$ and reached the periastron, which is below the disk.  After that, the
companion began to move away from RW~Aur\,A, crossed the plane $z=0$ once
more and now $(t=0)$ is above it.  Most likely, the circumstellar gas above
and below the disk plane experienced disturbances to various extent during
fly-by of the companion.  However, without making appropriate simulations,
it is difficult to say whether this circumstance has influenced the
distribution of matter {\it along the jet axis} to such an extent to
explain the observed asymmetry of jet's lobes.

 As we noted above, knots in CTTSs jets emerge due to the fact that
significant increase of gas outflow velocity occasionally occurs in the wind
formation region.  Apparently, such an episode, lasing for several days only
and accompanied by the stellar brightness increase by $\Delta B \approx 6$
mag, was observed in the star DF~Tau\,A \cite{Li-etal-01}.  In the last
column of Table~1, we presented estimates of the age of the spots $t_d$ (for
2017).  For seven spots, the times of their birth fall within the range of
the historical light curve of RW~Aur we constructed, however, we have not
found bursts comparable in scale with the burst of DF~Tau\,A near these
times.  On the other hand in the period of appearance of a giant "bubble"\,
in the jet of XZ~Tau\,A \cite{Krist-08}, no peculiarities in the light curve
of this star were noticed \cite{Dodin-etal-2016}.  Either flares of
RW~Aur\,A and XZ~Tau\,A have been missed because of their short duration, or
ejections of high velocity gas from CTTSs are not always accompanied by
flares of noticeable amplitude, what imposes restrictions on the mechanism
of ejections that is not known yet (see \cite{HHC-16,Frank-14} and
references therein).


\section*{Conclusion}

 We found from the comparison of RW~Aur\,A jet images separated by the time
interval of 21.3 yr that the most remoted from the star knots have emerged
$\approx 350$ years ago.  We suppose that at that moment the jet itself
emerged and the epoch of intensive accretion onto RW~Aur\,A has begun, caused
by the rearrangement of the structure of its protoplanetary disk due to
tidal influence of the companion RW~Aur~B.

  The companion, moving along a highly elongated orbit, passed through the
periastron $400-450$ years ago, having greatly changed the structure of the
outer disk regions of the primary.  But its direct influence on the inner
disk regions was apparently negligible.  Therefore, we assume that the
increase in accretion rate onto the primary is a reaction to a fundamental
change in conditions in the outer regions of the disk, which followed after
the sound wave, generated by these changes, passed along the disk in the
radial direction.

 It looks resonable to propose that the difference in the parameters of the
blue and red lobes of the jet is associated with the asymmetric distribution
of the circumstellar matter above and below the disk, which occurred due to
the tidal impact of the companion on the circumstellar gas.

  Finally, we also assume that deep, long-lasting dimmings of RW~Aur~A after
2010 due to its eclipse by the "dust screen"\, are also associated with the
rearrangement of the structure of the inner $(r < 1$ AU) disk regions, the
original cause of which is the close fly-by of the companion.  However, it
is still unknown not only what the nature of this screen is but also why it
has emerged only 350 yr after the beginning of the active accretion phase.

  Answers to the posed questions can be obtained from numerical
calculations considering the influence of the companion's fly-by on the
structure and dynamics of RW~Aur~A inner disk regions as well as the
distribution of the circumstellar matter around the primary.

\begin{figure}[ht]
 \begin{center} 
  \resizebox{16.0cm}{!}{\includegraphics{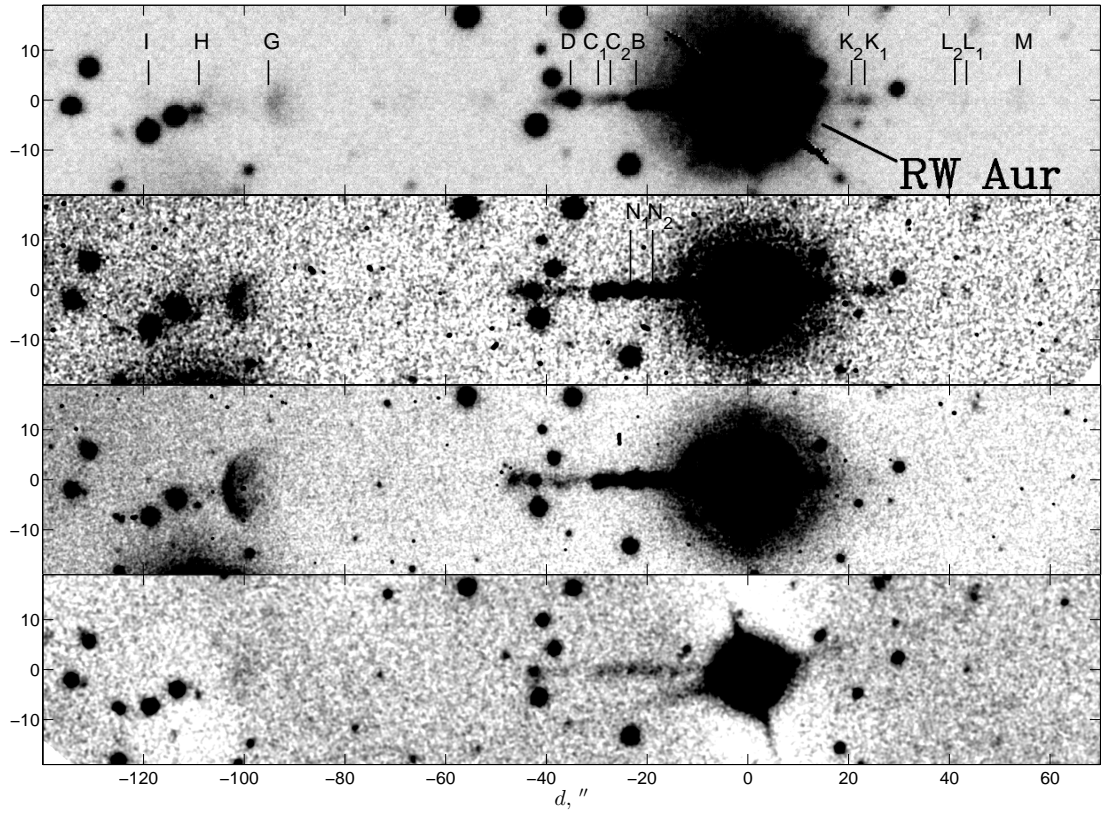}}
  \caption{ Images of the RW~Aur~A neighborhood obtained in different 
bands and epochs. The top panel -- 1995, in the [S II] band (adopted from
\cite{Mund-Eisloffel-98}); the second from the top -- 2017, in the [S II]
band; the third from the top -- 2017, in the $H_\alpha$ band; the bottom --
2017, in the [Fe II] 1.64 mkm band.  The jet axis (PA$\simeq 130^o$) in all
the images is directed along the horizontal axis, the coordinate center
coincides with the position of RW~Aur~A.}
\label{Fig-1}
  \end{center}
\end{figure}

  We did not compare intentionally in this paper the images of RW~Aur~A jet
observed in different bands. Such a comparison, supplemented by the
spectral data, is supposed to be done in the future with the goal of
studying physical conditions in previously unstudied regions of the jet at
distances $>30^{\prime\prime}.$ As we know up to now the images of these
regions in the $H_\alpha$ and [Fe II] $\lambda=1.64$ mkm bands have not even
been published, so we find it appropriate to present them in
Fig.\ref{Fig-1}.

\medskip
{\it {\bf Acknowledgments}}
\medskip

  We express our gratitude to the SAI team who carry out adjusting and
startup procedures at the Caucasus Mountain Observatory of the SAI MSU for
the opportunity to conduct observations, on which our study is based.  The
observed data on variable stars used in the paper are taken from the AAVSO
International Database being updated by observers from all over the world. 
In our work, we also used the data from the SIMBAD database supported by
CDS, Strasbourg, France.  The study of AVD (jet images reducing) and SAL
(formulation of the problem, interpretation of data) was conducted under the
financial support of the Russian Science Foundation Public Monitoring
Committee 17-12-01241.  Scientific equipment used in this study were bought
for the funds of the M.V. Lomonosov Moscow State University Program of
Development.


\newpage

\begin{table*}[]
\caption{Position and proper motion of knots in RW~Aur\,A jet} 
 \label{prop-motion-tab}
\medskip
\begin{tabular}{c|c|c|c|c|c|c}
\hline
Knot & $d_{1995},{}^{\prime\prime}$ & $d_{1997},{}^{\prime\prime}$ & 
$d_{2000}, {}^{\prime\prime}$  & $d_{2017},{}^{\prime\prime}$ &
$\mu$,${}^{\prime\prime}$/yr & $t_d$, yr \\
\hline
$M$   & 54.0 &      &        &      &       &     \\
$L_1$ & 43.4 &      &        &      &       &     \\
$L_2$ & 41.1 &      &        &      &       &     \\
$K_1$ & 23.2 &      &        & 26.5 & 0.154 & 170 \\
$K_2$ & 20.6 &      &        & 24.2 & 0.170 & 140 \\
$R_1$ &      & 11.1 &        & 14.0 & 0.153 &  92 \\
$R_2$ &      & 7.9  &        & 12.3 & 0.231 &  53 \\
$R_4$ &      & 2.92 & 3.45   & 6.8  & 0.205 &  33 \\
$R_?$ &      &      & 0.25   & 2.6  & 0.145 &  18 \\
\hline
$B_3$ &          &         & $-1.25$   & $-5.4$    & $-0.258$ &  21 \\
$N_3$ &          &         &           & $-14.2$   &          &     \\
$N_2$ &          &         &           & $-18.9$   &          &     \\
$N_1$ &          &         &           & $-23.3$   &          &     \\
$B$   & $-22.2$  &         &           & $-28.7$   & $-0.305$ &  94 \\
$C_2$ & $-27.3$  &         &           & $-34.7$   & $-0.347$ & 100 \\
$C_1$ & $-29.7$  &         &           & $-37.2$   & $-0.359$ & 104 \\
$D$   & $-35.2$  &         &           & $-42.5$   & $-0.341$ & 125 \\
$G$   & $-95.2$  &         &           & $-103.2$  & $-0.372$ & 280 \\
$H$   & $-109.1$ &         &           &           &          &     \\
$I$   & $-119.0$ &         &           &           &          &     \\
\hline                                                                             
\end{tabular}
\end{table*}

\newpage

\newpage

\begin{figure}[ht]
 \begin{center} 
  \resizebox{14.0cm}{!}{\includegraphics{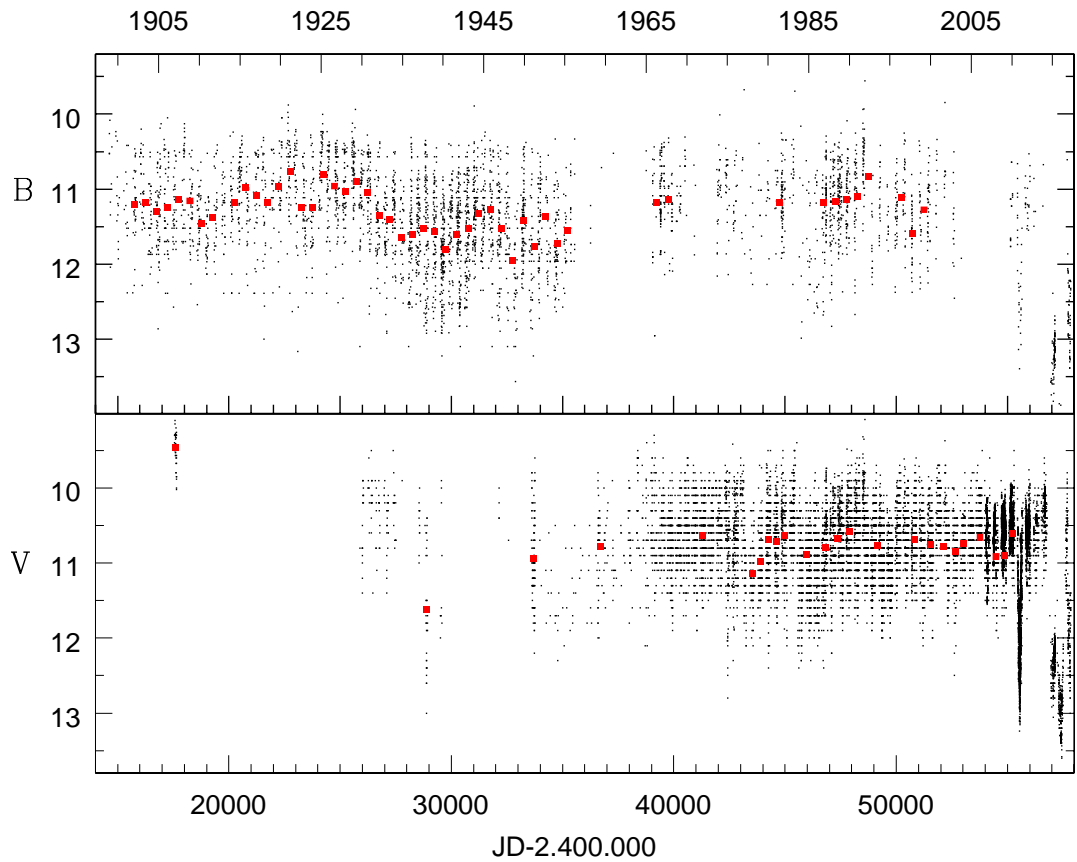}}
  \caption{ Historical light curve of RW Aur in the B (top panel) and V 
(bottom panel) bands. Red squares denote average brightness for seasons 
before 2010 yr with more than 30 measurements.}
\label{Fig-2}
  \end{center}
\end{figure}

\newpage

\begin{figure}[ht]
 \begin{center}
  \resizebox{14.0cm}{!}{\includegraphics{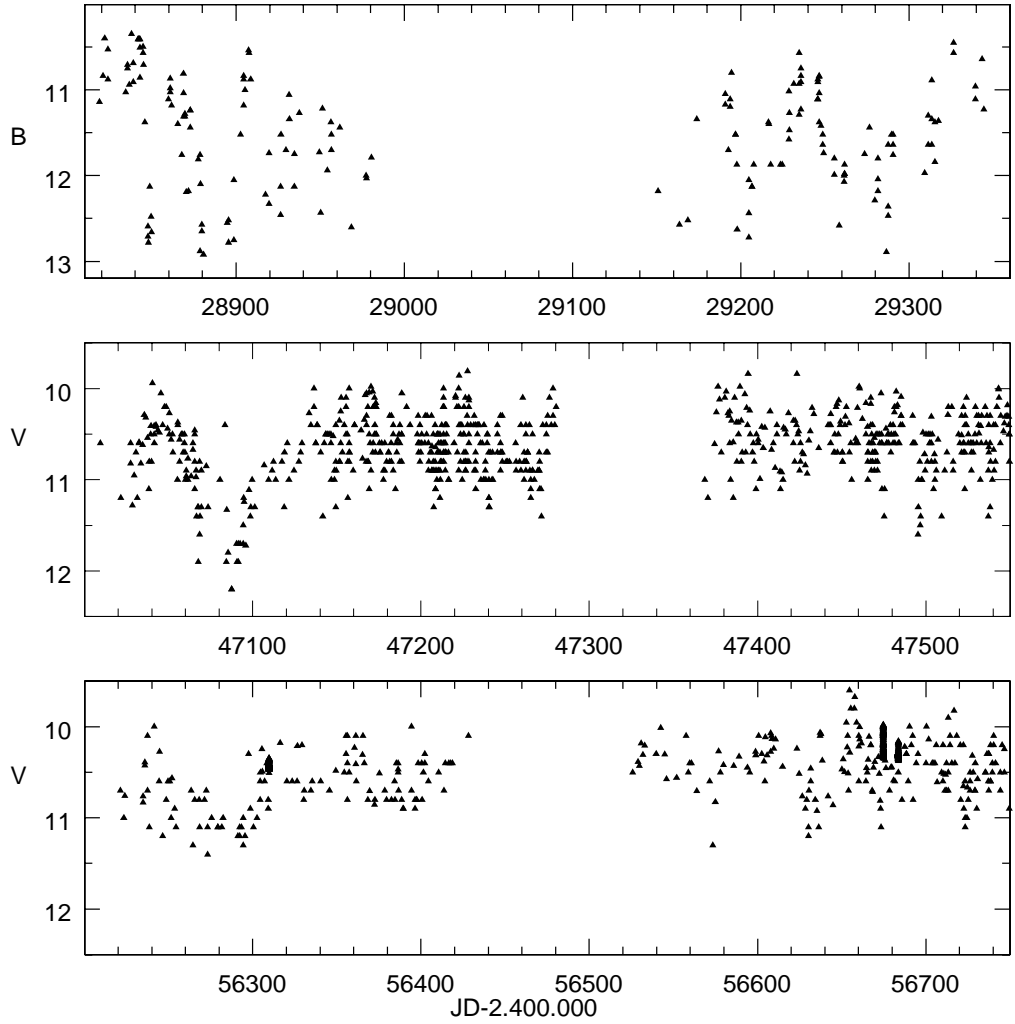}}
  \caption{Selected parts of RW Aur's historical light curve showing episodes 
of dimmings happened before 2010 yr.}
\label{Fig-3}
  \end{center}
\end{figure}

\end{document}